\documentclass[apj]{emulateapj}
\usepackage{graphicx,txfonts,amssymb,amsfonts,natbib,times,hyperref}

\begin{document}
\submitted{Accepted to The Astrophysical Journal}

\newcommand{\boot}{Bo\"otes}
\newcommand{\kms}{~km~s$^{-1}$}
\newcommand{\logh}{$+5\log h_{70}$}
\newcommand{\ho}{~}
\newcommand{\hi}{$h^{-1}_{70}$~}
\newcommand{\msun}{M$_\odot$}
\newcommand{\degtwo}{deg$^2$}
\newcommand \sbu {mag arcsec$^{-2}$}
\newcommand \firstpaper {(Paper I)}
\newcommand{\spitzer}{\textit{Spitzer}}
\newcommand{\galex}{{\it GALEX}}
\newcommand{\herschel}{\textit{Herschel}}
\newcommand{\hst}{\textit{HST}}
\newcommand{\chandra}{\textit{Chandra}}
\newcommand{\irac}{IRAC}
\newcommand{\irs}{IRS}
\newcommand{\mips}{MIPS}
\newcommand{\wfc}{WFC3}
\newcommand{\acs}{ACS}
\newcommand{\kcorrect}{\texttt{kcorrect}}
\newcommand{\galfit}{\texttt{GALFIT}}
\newcommand{\multidrizzle}{\texttt{MultiDrizzle}}
\newcommand{\hyperz}{\texttt{HyperZ}}
\newcommand{\ergflux}{~erg~s$^{-1}$~cm$^{-2}$}
\newcommand{\micr}{$\mu$m}
\newcommand{\mfive}{M$_{500}$}
\newcommand{\mtwo}{M$_{200}$}
\newcommand{\rfive}{$r_{500}$}
\newcommand{\rtwo}{$r_{200}$}
\newcommand{\target}{{IDCS J1426.5+3508}}
\newcommand{\bootarea}{{8.82}}
\newcommand{\sz}{Sunyaev-Zel'Dovich}
\newcommand{\lcdm}{$\Lambda$CDM}
\renewcommand{\d}{\mathrm{d}}

\title{IDCS J1426.5+3508: Cosmological Implications of a Massive,
  Strong Lensing Cluster at $z=1.75$} \shorttitle{A Strong Lensing
  Cluster at $z=1.75$}

\author{ Anthony H. Gonzalez\altaffilmark{1}, S. A.
  Stanford\altaffilmark{2,3}, Mark Brodwin\altaffilmark{4,5},
  Cosimo Fedeli\altaffilmark{1}, \\
  Arjun Dey\altaffilmark{6}, Peter R. M. Eisenhardt\altaffilmark{7},
  Conor Mancone\altaffilmark{1}, Daniel Stern\altaffilmark{7}, Greg
  Zeimann\altaffilmark{2} }

\altaffiltext{1}{Department of Astronomy, University of Florida, Gainesville, FL 32611-2055}
\altaffiltext{2}{Department of Physics, University of California, One Shields Avenue, Davis, CA 95616}
\altaffiltext{3}{Institute of Geophysics and Planetary Physics, Lawrence Livermore National Laboratory, Livermore, CA 94550}
\altaffiltext{4}{Department of Physics and Astronomy, University of Missouri, 5110 Rockhill Road, Kansas City, MO, 64110}
\altaffiltext{5}{Harvard-Smithsonian Center for Astrophysics, 60 Garden Street, Cambridge, MA 02138}
\altaffiltext{6}{NOAO, 950 North Cherry Avenue, Tucson, AZ 85719}
\altaffiltext{7}{Jet Propulsion Laboratory, California Institute of Technology, Pasadena, CA 91109}

\begin{abstract}
  The galaxy cluster \target\ at $z=1.75$ is the most massive galaxy
  cluster yet discovered at $z>1.4$ and the first cluster at this
  epoch for which the Sunyaev-Zel'Dovich effect has been observed.  In
  this paper we report on the discovery with \hst\ imaging of a giant
  arc associated with this cluster. The curvature of the arc suggests
  that the lensing mass is nearly coincident with the brightest
  cluster galaxy, and the color is consistent with the arc being a
  star-forming galaxy. We compare the constraint on \mtwo\ based upon
  strong lensing with Sunyaev-Zel'Dovich results, finding that the two
  are consistent if the redshift of the arc is $z\ga3$.  Finally, we
  explore the cosmological implications of this system, considering
  the likelihood of the existence of a strongly lensing galaxy cluster
  at this epoch in a \lcdm\ universe. While the existence of the
  cluster itself can potentially be accomodated if one considers the
  entire volume covered at this redshift by all current high-redshift
  cluster surveys, the existence of this strongly lensed galaxy
  greatly exacerbates the long-standing giant arc problem. For
  standard \lcdm\ structure formation and observed background field
  galaxy counts this lens system {\it should not exist}.
  Specifically, there should be {\it no} giant arcs in the entire sky
  as bright in F814W as the observed arc for clusters at $z\ge1.75$,
  and only $\sim 0.3$ as bright in F160W as the observed arc.  If we
  relax the redshift constraint to consider all clusters at $z\ge1.5$,
  the expected number of giant arcs rises to $\sim 15$ in F160W, but
  the number of giant arcs of this brightness in F814W remains zero.
  These arc statistic results are independent of the mass of
  \target. We consider possible explanations for this discrepancy.
\end{abstract}


\section{Introduction}

Galaxy clusters have historically played a central role in cosmology,
with the most massive and distant systems providing the most profound
insights. For example, observations of the Coma cluster provided the
first evidence for dark matter \citep{zwicky1933}, while the existence
of exceptionally massive clusters at early times was an important
argument for $\Omega_0<1$ \citep[e.g.,][and references
therein]{carlberg1996,donahue1998}. In recent years much attention has
been given to the question of whether the most distant, highest mass
clusters are consistent with a standard Gaussian \lcdm\ cosmology, or
whether one must invoke non-Gaussianity of the initial density
fluctuations from inflation that seed structure formation.  While
these analyses have yielded divergent results
\citep{hoyle2011,enqvist2011,cayon2011,williamson2011}, it is clear
that the most massive, distant clusters remain valuable cosmological
probes.

The galaxy cluster that is the focus of this paper was detected as
part of the IRAC Distant Cluster Survey (IDCS), an ongoing \bootarea\
\degtwo\ survey within the \spitzer\ Deep, Wide-Field Survey
\citep[SDWFS][]{ashby2009} region that employs full photometric
redshift probability distributions for a 4.5$\mu$m-selected galaxy
catalog to identify galaxy clusters at $0<z<2$.  This program extends
the IRAC Shallow Cluster Survey (ISCS; Eisenhardt et al. 2008), which
has yielded the largest sample of spectroscopically confirmed clusters
at $1<z<1.5$ by pushing both to lower mass and higher redshift. The
cluster \target\ was identified as a strong candidate for a high mass,
$z>1.5$ cluster in this program and targeted for detailed
follow-up. Spectroscopic observations with the \hst\ \wfc\ grism and
LRIS on Keck, described in detail in \citet{stanford2011}, confirm
that this cluster lies at $z=1.75$.

While the existence of a cluster at this redshift is not surprising,
multiple lines of evidence now suggest that this is a truly massive
cluster. \citet{stanford2011} report a \chandra\ X-ray mass estimate
of \mtwo $\simeq5.6\times10^{14}$ \msun, while Sunyaev-Zel'Dovich
imaging from \citet{brodwin2011} implies that the mass contained
within a region overdense by a factor of 500 relative to critical
density is \mfive $=2.6\pm 0.7 \times10^{14}$ \msun, (\mtwo$
\simeq4.2\times10^{14}$ \msun\ for a typical halo concentration).  For
comparison, this Sunyaev-Zel'Dovich mass is only $\sim40$\% lower than
that of XMMU J2235.3-2557 at $z=1.39$ \citep{mullis2005,rosati2009},
which is the only published cluster at $z\ga1.2$ more massive than
\target.  Meanwhile, the one spectroscopically confirmed galaxy
cluster at higher redshift \citep[$z=2.07$,][]{gobat2011} has an
estimated total mass of $5.3-8\times10^{13}$ \msun\ -- a factor of
five to ten lower than \target.

In this paper we focus upon the discovery of a giant arc associated
with this cluster and the implications of its existence in the context
of \lcdm\ structure formation. The layout of this paper is as
follows. In \S \ref{sec:arcimage} and \ref{sec:arcspec} we present the
discovery and attempted spectroscopy of the giant arc. In \S
\ref{sec:mass} we derive strong lensing constraints on the cluster
mass, and discuss the redshift regime over which these constraints are
consistent with the SZ mass.  We then extend our discussion in \S
\ref{sec:arcstats} to consider the probability for the discovery of a
giant arc associated with this cluster. Finally, in \S
\ref{sec:summary} we summarize our results and consider potential
theoretical modifications that may resolve the arc statistic
discrepancy. Throughout this paper we use cosmological parameters
consistent with the seven year WMAP results
\citep[$\Omega_\Lambda=0.728$, $H_0=70.4$ \kms,
$\sigma_8=0.809$;][]{komatsu2011}.

\section{Detection of a Giant Arc}
\label{sec:arcimage}

We obtained \hst\ imaging with the Advanced Camera for Surveys
\citep[\acs;][]{Ford2003} and Wide-Field Camera 3
\citep[\wfc;][]{kimble2008} as part of Cycle 17 program 11663 (PI:
Brodwin) between 08 July 2010 and 07 November 2010.  The total
exposure times in F814W and F160W were 4.5 ks and 2.6 ks,
respectively. Further details are provided in
\citet{stanford2011}. Subsequent grism observations in Cycle 18,
coupled with Keck spectroscopy, confirm that the galaxy cluster is
real and at $z=1.75$ \citep{stanford2011}.

Within the \hst/\acs\ and \wfc\ imaging we identify a highly elongated
object which we interpret as a strong arc lensed by the cluster.  We
present a composite F814W+F160W image of the cluster field in
Fig. \ref{fig:ims}, highlighting this object. The length is 4$\farcs$8
but the width is unresolved in the \hst\ photometry, and hence it
easily satisfies the standard length-to-width criteria for a giant arc
\citep[$l/w>10$;][]{wu1993}. The curvature of the object is consistent
with lensing by the cluster potential, and the color can be used to
further constrain the nature of the object.

\subsection{Arc Photometry}

\begin{figure*}
\epsscale{0.49}
\epsscale{0.455}
\plotone{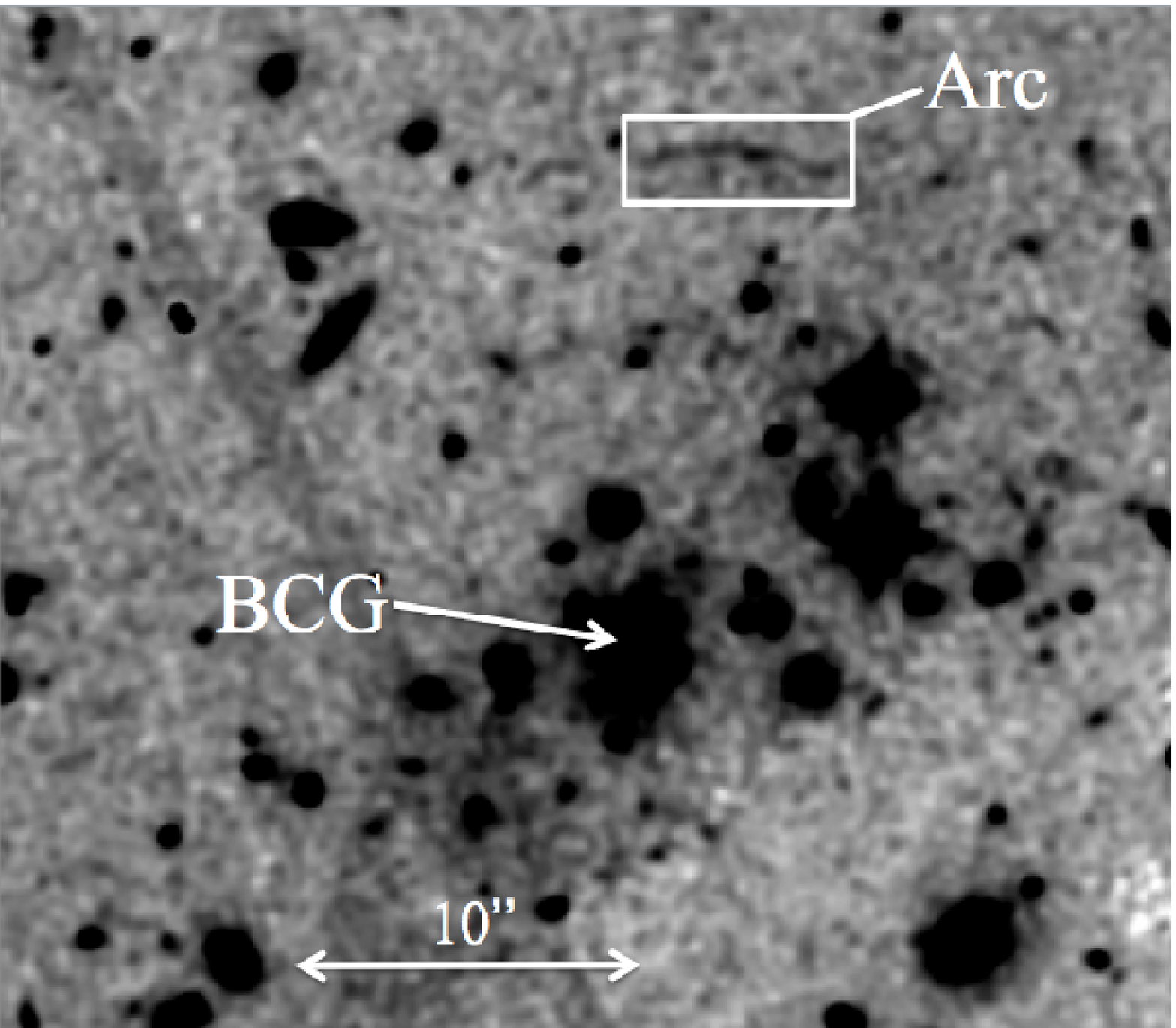}
\epsscale{0.455}
\plotone{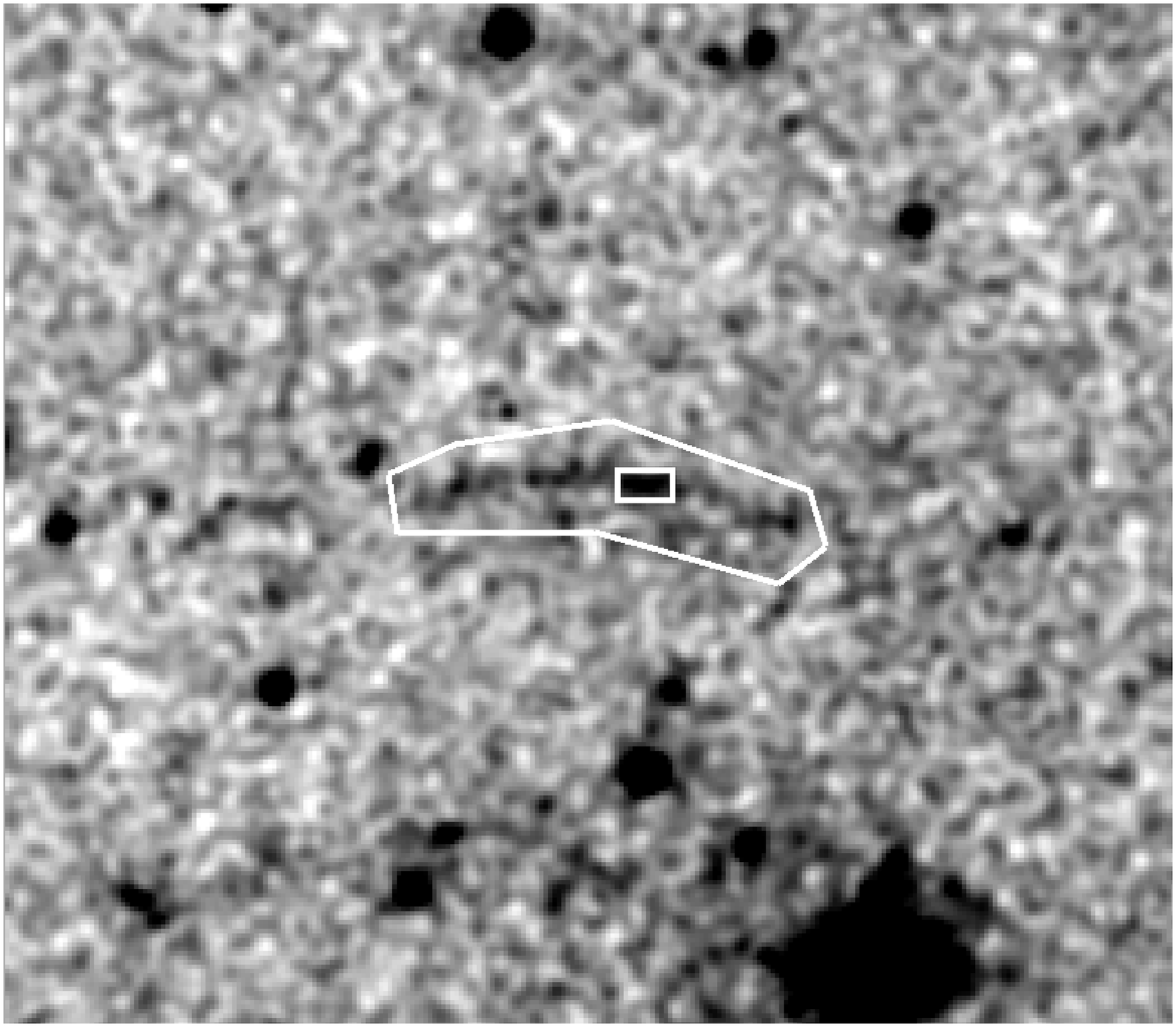}
\epsscale{1.0}
\caption{ {\it Left--} Combined F814W+F160W image of the cluster
  center and giant arc. The field of view is 30$\arcsec$; North is up
  and East is to the left. {\it Right--} Zoomed-in version of the same
  image centered on the arc. The polygon is the aperture used to
  extract the arc photometry, while the smaller rectangle is the
  region within which the color was determined. The field of view is
  12$\farcs$5 across. The images have been smoothed with 5 pixel and 3
  pixel Gaussian kernels, respectively, to enhance the
  contrast.}\label{fig:ims}
\end{figure*}

We extract the magnitude of the arc within a polygonal aperture
constructed to enclose the full extent of the arc. The enclosed area
within the aperture is 9.0 arcsec$^2$, and the extent along the major
axis is 6$\farcs$7 (Fig. \ref{fig:ims}). As part of this procedure we
first use Source Extractor \citep{bertin1996} to generate a background
map.  The flux and background are measured both within the source
aperture and for an ensemble of blank-sky apertures surrounding the
arc, from which we can calculate the aperture-to-aperture photometric
variance.  We measure F814W$=24.29\pm0.31$ mag and
F160W$=23.75\pm0.21$ mag (AB).  Given the large uncertainties in the
total magnitude, the integrated color of the arc,
F814W$-$F160W$=0.55\pm0.37$ mag, is only modestly constrained.  To
obtain an improved estimate of the color, we 
recompute the color within a smaller, $0\farcs8\times0\farcs4$
rectangular aperture that includes the region with the highest
signal-to-noise ratio (Fig. \ref{fig:ims}). Within this aperture we
obtain a more precise, statistically consistent color,
F814W$-$F160W$=0.25\pm0.13$ mag.  As expected for an arc, we see no
evidence for color changes along its length that might be indicative
of a chance superposition of sources.

What does this color imply about the source galaxy? The fact that the
source is not a drop-out in F814W constrains the redshift to be
$z\la6$. The observed color is similar to that of $z\sim4$ $B-$band
dropouts \citep{vgonzalez2011}. For a somewhat more complete picture
of plausible redshifts, we consider the predicted colors from a suite
of \citet{bruzual2003} stellar population models.  For a
\citet{chabrier2003} mass function and solar metallicity, the observed
color is inconsistent with passively evolving stellar populations at
all redshifts. To be precise, single burst models in which star
formation has ceased more than $\sim$100 Myrs earlier uniformly
predict F814$-$F160W colors that are significantly redder than
observed regardless of the formation redshift.  Conversely, models
with an exponentially declining star-formation rate ($\tau-$models)
can successfully reproduce the observed color at higher
redshifts. These models place a lower bound on the redshift as a
function of $\tau$ and the assumed formation redshift.  For $\tau=1$
Gyr, $1\sigma$ consistency with the observed color requires $z>1.75$
(2.5) for $z_{\mathrm{form}}>2$ (3).  Smaller values of $\tau$
increase the minimum redshift.

\subsection{Arc Geometry}
The location of the arc is such that no nearby individual galaxies
appear to be contributing significantly to the lensing, and the
curvature is consistent with the centroid being nearly coincident with
the brightest cluster galaxy (BCG).  If we make the assumption that
the lensing is indeed centered upon the BCG, then the radius of the
arc is $\theta=14\farcs6\pm0\farcs2$. There is however no guarantee
that the BCG lies directly at the bottom of the potential in the
absence of a detailed lensing model, so for the analysis below we
assume an uncertainty of 30 kpc in the centroid, which corresponds to
a 3$\farcs4$ uncertainty in the radius of the arc. For local clusters
from the LoCuSS program roughly 68\% of clusters have offsets between
the BCG and X-ray peak less than this value \citep{sanderson2009}.

\section{Gemini and \hst\ Spectroscopy}
\label{sec:arcspec}

We attempted to obtain a redshift for the giant arc using Gemini
North.  We targeted Ly$\alpha$ emission during 6.5~hr of Director's
Discretionary Time, using GMOS for long slit spectroscopy.  The data
were acquired between 26 June 2011 and 06 July 2011 (UT) using the
B600 grism, which is blazed at 4610 $\AA$.  We observed with a central
wavelength of 5150 $\AA$\ and two pixel binning in both the spatial
and spectral directions using a 1\farcs0 slit.  The resultant spatial
and spectral resolutions are 0\farcs146 pix$^{-1}$ and 0.9 $\AA$\
pix$^{-1}$, respectively, and the slit was positioned to lie along the
long axis of the arc at a position angle of 270.85 degrees.  We
obtained 13 $\times$ 1780~s exposures, dithered in both the spatial
and the spectral dimensions, for a total on-source exposure time of
6.43~hr.  The seeing during these observations ranged from 0\farcs67
to 1\farcs08, with a median seeing of 0\farcs77 based on an early
M-type star which was serendipitously observed in the majority of the
observations.  Conditions during the observations were mostly
photometric, but included some data taken during 70th percentile
(patchy cloud) conditions.

The redshift constraints arising from the cluster redshift (i.e., $z >
1.75$) and the fact that the arc is detected in the F814W imaging
(i.e., $z \la 6$) provide a first bound on the redshift.  Within this
range we focused upon $z \la 4$ and designed the observations to be
sensitive to strong Ly$\alpha$ emission (comparable to a Lyman Alpha
Emitter), if present for $2 \la z \la 4$.  We reduced the data using
standard long slit procedures within IRAF.  Unfortunately, we detect
neither any continuum nor any emission lines at 3590 $\AA < \lambda <$
6660 $\AA$ ($2.0 < z < 4.5$ for Ly$\alpha$).

We also attempted to obtain a redshift for the arc using \hst/\wfc\
grism data from program 12203 (PI Stanford). The data and reduction
procedure are both described in \citet{stanford2011}. Neither
continuum nor emission lines for the arc were detected with either the
G102 or G141 grisms, which together cover the wavelength range
$0.8$\micr\ $< z <1.65$\micr. Using simulations run with the aXeSIM
software, we calculate that the corresponding emission line detection
threshold corresponds to $f<4\times10^{-16}$ \ergflux\ at 0.95\micr\
and $f<3\times10^{-16}$ \ergflux\ at 1.35\micr\ (5
$\sigma$).{\footnote{\url{http://axe.stsci.edu/axesim/}}} Given the
non-detection of Ly$\alpha$ with either GMOS or \hst/\wfc\
spectroscopy, \hst\ narrow- or medium-band imaging may be the most
promising avenue for refining the redshift estimate for this arc.

\section{Cluster Mass from Strong Lensing}
\label{sec:mass}

\subsection{Enclosed Mass within the Arc}

Under the assumption of circular symmetry for the cluster lens, we
calculate the total mass enclosed by the giant arc as a function of
the source redshift. In this case the arc radius $\theta_\mathrm{a}$
($\sim 125$ kpc at $z_\mathrm{L} = 1.75$) identifies the radius of the
tangential critical curve, which can be easily related to the enclosed
mass through the relation,

\begin{equation}
M_\mathrm{a} = \pi\Sigma_\mathrm{c}~\theta^2_\mathrm{a},
\end{equation}
where $\Sigma_\mathrm{c}$ is the lensing surface critical density, which reads

\begin{equation}
  \Sigma_\mathrm{c} = \frac{c^2}{4\pi G} \frac{D_\mathrm{S}}{D_\mathrm{L}D_\mathrm{LS}}.
\end{equation}
In this equation $D_\mathrm{L}$, $D_\mathrm{S}$, and $D_\mathrm{LS}$
are the angular diameter distances to the lens, to the source, and
from the lens to the source, respectively.{\footnote{We refer the
    reader to \cite{ksw2006} for a detailed review of gravitational
    lensing.}}

We emphasize that this enclosed mass is independent of the specific
density profile assumed for the lens.  One important caveat in this
estimate, however, is that the assumption of circular symmetry is
known to yield an overestimate of the enclosed mass for more realistic
systems with intrinsic ellipticity. To approximately account for this
effect, we assume that the circular model results in a factor of
$\sim1.6$ overestimate of the mass, consistent with
\citet{bartelmann1995}, and quote values below that include this
correction.

In Figure \ref{fig:lensingmass} we show the resulting enclosed mass as
a function of the source redshift. The closer the source is to the
deflector, the larger the enclosed mass needs to be due to the
geometric suppression of the lensing efficiency.  The uncertainty in
the enclosed mass shown in the Figure corresponds to the uncertainty
in the arc radius, for which we adopt the nominal value of $30$ kpc
(see \S \ref{sec:arcimage}). The value for the enclosed mass reaches a
lower limit of M$_{a}=6.9\pm 0.3\times 10^{13}$ \msun\ for $z_s=6$.
The enclosed mass in this central 125 kpc region, which contains
minimal assumptions, already is comparable to the total mass inferred
for the only spectroscopically confirmed cluster at higher redshift
\citep{gobat2011}.

\subsection{\mtwo}
\label{sec:mtwo}

The next step is to estimate the total mass within \rtwo\ for the
cluster. This problem is underconstrained, necessitating several
simplifying assumptions. We initially assume that the density profile
of the cluster is well represented by a spherical NFW model
\citep{nfw1996,nfw1997}. For a given virial mass, we compute the
concentration of the dark matter halo according to the prescription of
\citet{gao2008}, which is in turn a modified version of the original
NFW prescription. The \citet{gao2008} formula has been extensively
tested against numerical simulations, including the high redshift
regime relevant to the current analysis, and is expected to provide
improvement over the prescriptions of \citet{eke2001} and
\citet{bullock2001}.

To account for asymmetries in the cluster mass distribution, we next
assign a non-vanishing ellipticity to the lensing potential, according
to the procedure summarized in 
\citet{meneghetti2003}. Finally, we assume that the arc is produced by
a source lying near one of the caustic cusps situated along the major
axis of the lens, so that the arc radius corresponds to the maximum
elongation of the critical curve. Thus, we vary \mtwo\ until we find a
match between this maximum elongation and $\theta_\mathrm{a}$. 

\begin{figure}
\plotone{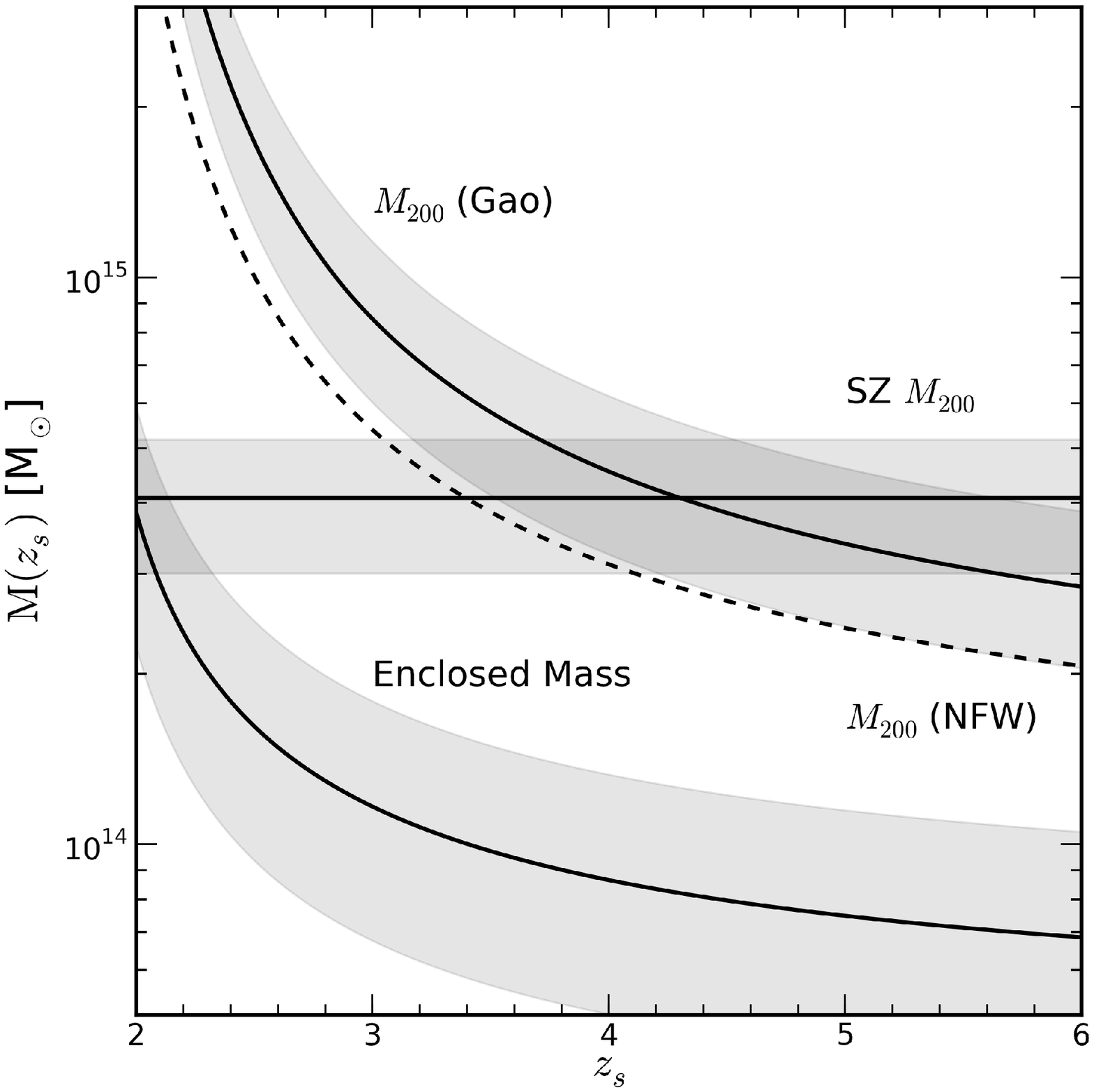}
\caption{Mass of \target\ as a function of redshift of the lensed
  source. The lower curve corresponds to the mass enclosed within the
  arc, with the shaded region denoting the uncertainty associated with
  the offset of the BCG relative to the cluster potential. The upper
  curve is the inferred \mtwo\ assuming the \citet{gao2008}
  prescription for the concentration and an ellipticity
  $e_\mathrm{m}\simeq 0.32$ for the cluster dark matter halo. In this
  case the uncertainty denoted by the shaded region is dominated by
  the intrinsic scatter in the distribution of halo ellipticities. We
  also overplot as a dashed line the inferred mass if one instead uses
  the original NFW prescription for the halo concentration (which can
  be considered a lower bound). The horizontal line and associated
  uncertainties correspond to the \mtwo\ derived from
  Sunyaev-Zel'dovich observations. In this case the uncertainties do
  not include the potential systematic bias associated with
  extrapolating SZ scaling relations to higher redshift.}
\label{fig:lensingmass}
\end{figure}

In Figure \ref{fig:lensingmass} we show the resulting \mtwo\ as a
function of the source redshift.  We assume an ellipticity,
$e_\mathrm{m}\simeq 0.32$, consistent with the mean of the ellipticity
distribution presented in Figure 7 of \citet{fedeli2009}, and use the
standard deviation of this distribution $\sigma_e\simeq 0.074$ to
define the uncertainty shown by the shaded region.  A caveat to this
assumption is that this ellipticity distribution is derived at low
redshift. \citet{lee2005} however demonstrated that evolution of the
ellipticity distribution is expected to be negligible for $z<1.5$, and
sufficiently small at $z<2$ as to not impact our calculations.  From
this analysis the derived value for \mtwo\ approaches a lower limit of
\mtwo$=2.8^{+1.0}_{-0.4}\times 10^{14}$ \msun\ as $z_s\to 6$, where
the quoted uncertainty reflects the uncertainty in the ellipticity.
For reference, we also show the results obtained with the same
fiducial ellipticity $e_\mathrm{m}$, but assuming the original NFW
prescription for the concentration (dashed curve). Since at high
redshift the NFW concentration is always higher than other
prescriptions, the required \mtwo\ of the cluster is 35\%
smaller. Conversely, use of either the \citet{bullock2001} or
\citet{eke2001} prescriptions would lead to a larger value of \mtwo.

It should be noted that strong cluster lenses are usually a biased
subsample of the whole cluster population, in the sense that they tend
to be intrinsically more concentrated, and to be prolated with the
major axis aligned along the line of sight
\citep{hennawi2007,meneghetti2010}. This bias is expected to be even
more severe in redshift and/or mass ranges where strong lensing is
particularly rare, such as the case under consideration.  It has also
recently been observed that there may be a stronger than expected
correlation between concentration and cluster mass, with lower mass
systems having higher than expected concentrations
\citep[e.g.,][]{schmidt2007,ettori2010,oguri2011}.  Therefore, a
concentration higher than that provided by the \citet{gao2008}
prescription, which is a mean over the entire cluster population,
might actually be more realistic in this circumstance. As an example,
for a sample of lensing clusters at $z\sim0.5$ \citet{oguri2009} found
concentrations a factor of $\sim2$ higher than would be predicted by
the \citet{gao2008} distribution.  Simulations by
\citet{meneghetti2010} also indicate that projection effects may yield
observed concentrations that are enhanced by up to a factor of two. In
the current case, if we assume that the concentration is a factor of
two above the \citet{gao2008} prescription, then the derived \mtwo\
would decrease by roughly a factor of 2.2.

In Figure \ref{fig:lensingmass} and Table \ref{tab:masses} we also
present the masses and associated uncertainties derived from the \sz\
analysis \citep{brodwin2011}.  The two approaches appear to yield
consistent estimates for \mtwo\ if the source redshift lies at
$z\ga3.5$, or more conservatively $z\ga3$ if one includes the
potential for reducing the lensing mass by up to a factor of $\sim$2
if the halo concentration is larger than for a typical
cluster. Coupled with the color of the arc, these factors together
argue that the most plausible redshift is $3\la z\la 6$.

\section{Cosmological Implications}
\label{sec:arcstats}

The redshift of this cluster makes it a unique and interesting test
for cosmological structure formation.  To be specific, the most
distant clusters known to host giant arcs prior to this study lie at
$z\sim1$ \citep[e.g.,][]{gladders2003,huang2009}.  \target\
significantly extends the redshift baseline over which arcs are known
to exist. In this section we consider the probability for this
massive, strong lensing cluster to exist and be detected in our
survey.  Specifically, given a standard \lcdm\ cosmology with the
seven year WMAP cosmological parameters, what is the probability of
detecting a giant arc of this brightness behind a cluster at $z>1.75$?

In order to estimate how rare the observed gravitational arc is, we
evaluate the number of arcs in the whole sky that are expected to be
produced by galaxy clusters at redshift larger than $z_\mathrm{L} =
1.75$.  We first estimate, for a fixed source redshift $z_\mathrm{S}$,
the contribution to the optical depth by structures in the desired
redshift range,

\begin{equation}
  \tau_q(z_\mathrm{S}) = \frac{1}{4\pi D_\mathrm{S}}\int_{z_\mathrm{L}}^{z_\mathrm{S}} \d z\int_0^{+\infty} \d M~n(M,z)\left| \frac{dV(z)}{dz} \right|\sigma_q(M,z).
\end{equation}
In the previous equation $n(M,z)$ is the mass function of cosmic
structures, $dV(z)/dz$ represents the comoving volume of space per
unit redshift, and $\sigma_q(M,z)$ stands for the cross section of
individual clusters for images having the morphological property
$q$. In what follows we assume that $q$ is a length-to-width ratio
$\ge 10$, as customary in arc statistics studies, and employ the
\citet{tinker2008} mass function.

The total number of arcs with the property $q$ that are observed in
the sky with a magnitude brighter than $m$ then simply reads

\begin{equation}
  N_q(m) = 4\pi~n_\mathrm{S}(m)\int_{z_\mathrm{L}}^{+\infty} p(z_\mathrm{S},m)\tau_q(z_\mathrm{S})\d z_\mathrm{S},
\label{eqn:narcs}
\end{equation}
where $p(z_\mathrm{S},m)$ is the source redshift distribution, while
$n_\mathrm{S}(m)$ represents the observed number density of sources
with magnitude lower than $m$, i.e., the cumulative number counts. We
adopt the redshift distribution and number counts for sources in the
{\it Hubble} UDF provided by \citet{coe2006} for F775W and F160W, and
correct the number counts for the lensing magnification bias using the
same procedure detailed in \citet{fedeli2008} and \citet{fedeli2009}.

\begin{figure}
\plotone{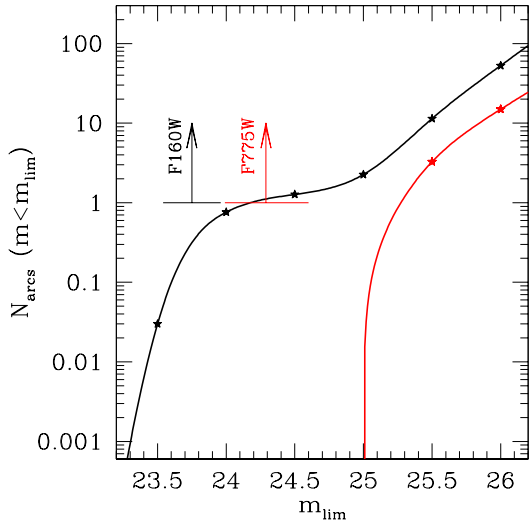}
\caption{Predicted number of giant arcs over the entire sky as a function of magnitude in F775W and F160W for clusters at $z>1.75$.
The points correspond to the results from our simulations, while the curves are spline interpolations between the data points. The arrows represent the all-sky lower limits derived from the observed arc in \target, with the width at the bottom of the arrows corresponding to the photometric uncertainty. We note that finding one arc per \bootarea\ \degtwo\ would correspond to $\sim 4700$ arcs all-sky. 
\label{fig:counts}}
\end{figure}

For the practical computation of the optical depth, we use the
strategy detailed in \citet{fedeli2008} and references
therein. Briefly, merger trees are constructed based on the extended
\citet{press1974} theory, which represent a model of the cluster
population. A lensing potential ellipticity is assigned to each
cluster, extracted from the distribution shown in \citet{fedeli2009},
and cluster dynamical activity occurring at the knots of the merger
trees is suitably modeled.  Individual clusters are modeled as NFW
profiles with the assigned potential ellipticity, and the
concentration is linked to the mass through the \citet{gao2008}
prescription for the concentration, consistent with \S \ref{sec:mtwo}.
We include a lognormal scatter in the concentration with
$\sigma_c=0.2$, as in \citet{fedeli2007}.  Finally, the cluster cross
sections for giant arcs are computed using the fast and semi-analytic
prescription of \citet{fedeli2006}, and the optical depth integrals
are approximated by using a Monte-Carlo scheme.

Figure \ref{fig:counts} shows the resulting number of arcs expected
across the full sky as a function of magnitude in F775W and F160W. For
our observed arc we assume a color correction F775W$-$F814W$\approx
0.0$ (AB), consistent with the expected color of a star-forming galaxy
at this epoch.  For the observed magnitudes we expect to find {\bf no}
arcs over the entire sky as bright in F814W as the one we observe and
only $\sim 0.3$ as bright in F160W.  Indeed, none are expected within
0.5 magnitudes of the brightness of the arc in F814W.  Given that the
area of our survey is only \bootarea\ \degtwo\ ($2\times10^{-4}$ of
the full sky), the detection of an arc is highly implausible.  For
reference, we test the sensitivity of this result to the redshift of
the lensing cluster. Even if one were to consider the entire cluster
population at $z>1.5$ rather than setting the observed cluster
redshift as the minimum permitted redshift, then the expected number
of arcs detected remains zero in F814W and $\sim15$ all-sky in
F160W. The latter still leaves only a probability of $2\times10^{-3}$
that we would have detected such an arc in this survey.  Note that the
specific mass of \target\ does not enter into our calculations -- we
posed the question of how many arcs should be produced by {\it all}
clusters.

At lower redshifts the excess of giant arcs behind clusters relative
to predictions has been realized for well over a decade
\citep[e.g.,][]{bartelmann1998} and is known as the arc statistics
problem.  While significant effort has been devoted to reconcile the
observations with improved models that incorporate more detailed
physics and improved constraints on the source redshift distribution
\citep[e.g.,][]{bartelmann2003,wambsganss2004,dalal2004,li2006,fedeli2008,wambsganss2008,dalioso2011},
the issue is not wholly resolved \citep[e.g.,][]{meneghetti2011}.  The
current cluster greatly exacerbates the situation -- this arc simply
{\it should not exist}.

It is interesting to break down the above calculation of the number of
arcs (Equation \ref{eqn:narcs}) to ascertain the dominant factor that
drives the expected number of arcs to zero.  The two fundamental
physical quantities that determine the total number of arcs are the
total lensing optical depth of all clusters and the number density of
background sources to be lensed.

The lensing optical depth depends upon the products of the cluster
mass function and the lensing efficiency of individual clusters, which
is a strong function of cluster mass.  An obvious explanation would be
if \target\ exceeds the mass of clusters expected to exist in a \lcdm\
universe at this redshift. If lensing requires reaching a critical
mass threshold that is exceeded in reality but not in our model, then
the observed number of arcs would clearly exceed expectations.  In
that case, a plausible solution would be to invoke non-Gaussianity to
enhance the number of extremely massive clusters. It is however argued
in \citet{brodwin2011} that \target\ is consistent with \lcdm,
indicating that invoking non-Gaussianity is not necessarily the
appropriate solution.

A more subtle solution would be if the lensing cross-sections of
individual clusters systematically exceed the calculated theoretical
cross-sections. There is evidence that this lensing efficiency is
indeed systematically underestimated.  Several papers have argued that
including the impact of baryonic contraction can raise the lensing
cross-section by between 25\% and a factor of two
\citep{li2006,wambsganss2008}.  \citet{meneghetti2011} also find that
simulated clusters produce $\sim50$\% fewer arcs than X-ray clusters
at $z\approx0.5-0.7$.  While these corrections work in the proper
direction, it appears that even if one imposes reasonable physical
tweaks to account for the impact of baryonic infall on the density
profile, the change remains insufficient to account for the arc in
\target .  Moreover, as demonstrated by \citet{mead2010}, inclusion of
AGN feedback acts to counteract the impact of baryonic infall,
resulting in a smaller enhancement to the cross-section.  Even
ignoring the mitigating effect of feedback, the discrepancy is simply
too large.  Specifically, doubling the predicted number of arcs would
imply $\sim 1$ arc {\it all-sky} with the observed F160W magnitude,
and would still imply none with the observed F775W magnitude.

Another means of boosting the effective cross-section for an
individual system is via the presence of additional structures along
the line of sight. \citet{puchwein2009} used the Millenium simulations
to quantify the impact of such secondary structures. These authors
found that the typical impact is to enhance the cross section by
$10-25$\%, with enhancements of 50\% not uncommon for individual
systems. Again, this factor alone is insufficient to explain the
discrepancy.

The final factor that can drive the prediction of zero arcs is
underestimation of the surface density of galaxies sufficiently bright
to yield a source similar to the observed arc after magnification by
the cluster potential. One concern here is that we have used results
from the UDF to inform our redshift distribution for background
sources, yet this region is sufficiently small that cosmic variance is
a concern. As an alternate test, we also try an analytic prescription
for the background number counts based upon the $z\sim3$ luminosity
function of \citet{reddy2008} to model the background distribution. We
find however that this approach does not qualitatively alter the
results of our analysis. It therefore seems unlikely that error in the
source distribution is the origin of the discrepancy.

We therefore identify no obvious physical solution to explain the
existence of this arc, though exceptionally high concentrations seems
like the most promising avenue to explore. Finally, a last possible
solution would be if the observed source is not a background
arc. Given the combination of color, curvature, $l/w$, and lack of
color variation, this possibility also seems unlikely.

\begin{deluxetable}{lll}
\tablewidth{0pt}
\tablecaption{Derived Masses for \target}
\tablehead{
\colhead{Method} 	& \colhead{Radius} & \colhead{Mass ($10^{14}$ \msun)}   }
\startdata
Lensing	&  125 kpc &	$>0.69\pm 0.03$\\
Lensing	&  \rtwo\  &	$>2.8^{+1.0}_{-0.4}$\\
SZ     	&  \rfive\ &	$2.6\pm 0.7$\\
SZ     	&  \rtwo\  &	$4.1\pm 1.1$
\enddata
\tablecomments{The lensing limiting masses correspond to the maximum
  possible source redshift, $z_s=6$.  Both \mtwo\ masses presume a
  concentration consistent with the \citet{gao2008} relation. If the
  projected concentration is a factor of two higher then the lensing
  and SZ masses drop by factors of 2.1 and 1.2,
  respectively. 
  As noted in the text, \rfive\ and \rtwo\ are defined relative to
  critical density.  }
\label{tab:masses}
\end{deluxetable}

\section{Conclusions} \label{sec:discussion}
\label{sec:summary}

We have presented evidence for the existence of a giant arc behind the
massive galaxy cluster \target\ at $z=1.75$. This unique system
constitutes the highest redshift cluster known to host a giant arc.
From the strong lensing we derive an enclosed lensing mass within the
central 125 kpc of $M_\mathrm{a}>6.9\pm 0.3\times 10^{13}$ \msun,
which is comparable to (or exceeds) the {\it total} masses of most
other known clusters at $z>1.5$, confirming that \target\ is an
exceptionally massive cluster at this epoch.  Having derived the
enclosed mass, we next provide a comparison with the \sz-derived value
of \mtwo\ from \citet{brodwin2011}. We find that the two are
consistent if the arc lies at $z\ga3$, and derive a lower bound on the
mass of $M_{200}>2.8\times10^{14}$ \msun.

Finally, we investigate the cosmological implications of this system.
The greatest challenge posed by this cluster is explaining the
existence of the giant arc at all.  In \S \ref{sec:arcstats} we
demonstrate that under realistic assumptions for the lensing
cross-section, cluster mass function, and background galaxy
distribution, the total number of giant arcs behind clusters at
$z>1.75$ that are at least as bright as the observed arc is zero in
F814W and $<1$ in F160W.  Very simply, the arc we have discovered
behind \target\ is not predicted to exist.  If one considers an
ensemble of lensing clusters extending to somewhat lower redshift,
$z>1.5$, the tension with theoretical expectations decreases slightly
for F160W, but in F814W the number of predicted arcs remains zero.  We
briefly discuss possible explanations for this discrepancy, but find
no obvious solution. A tendency of higher concentrations in observed
clusters than simulated systems has the greatest potential to decrease
the disparity, but is unlikely to be sufficient to reconcile
theoretical models for arc statistics with the existence of a lensing
cluster at $z=1.75$.

Looking towards the future, a statistical sample of the most massive
clusters in the Universe at $z=1-2$ will provide the means with which
to ascertain the true frequency of arcs behind high-redshift
clusters. Much as the frequency of strong lensing clusters at
$z=0.5-1$ was a surprise a decade ago, it appears that this higher
redshift regime is poised to yield further unexpected discoveries.

\acknowledgements 
The authors thank the anonymous referee for suggestions that improved
the quality of this paper. We are also grateful to Gemini Observatory
for allocating Director's Discretationary time to obtain a redshift
for the giant arc. We appreciate the support provided by Nancy
Levenson and the rest of the Gemini staff for this project. Gemini
Observatory is operated by the Association of Universities for
Research in Astronomy, Inc., under a cooperative agreement with the
NSF on behalf of the Gemini partnership: the National Science
Foundation (United States), the Science and Technology Facilities
Council (United Kingdom), the National Research Council (Canada),
CONICYT (Chile), the Australian Research Council (Australia),
Minist\'{e}rio da Ci\^{e}ncia, Tecnologia e Inova\c{c}\~{a}o (Brazil)
and Ministerio de Ciencia, Tecnolog\'{i}a e Innovaci\'{o}n Productiva
(Argentina).  This work is based in part on data obtained at the
W.~M.~Keck Observatory, which is operated as a scientific partnership
among the California Institute of Technology, the University of
California and the National Aeronautics and Space Administration.
Support for this research was provided by NASA through HST GO programs
11663 and 12203.  AHG thanks Marusa Brada\v{c} for a useful discussion
about the arc, and also acknowledges support from the National Science
Foundation through grant AST-0708490.  The work by SAS at LLNL was
performed under the auspices of the U.~S.~Department of Energy under
Contract No. W-7405-ENG-48, and support for MB was provided by the
W.~M.~Keck Foundation.  The work of PRME and DS was carried out at Jet
Propulsion Laboratory, California Institute of Technology, under a
contract with NASA. The research activities of AD are supported by
NOAO, which is operated by the Association of Universities for
Research in Astronomy (AURA) under a cooperative agreement with the
National Science Foundation.

\bibliographystyle{apj}
\bibliography{ms}

\begin{thebibliography}{57}
\expandafter\ifx\csname natexlab\endcsname\relax\def\natexlab#1{#1}\fi

\bibitem[{{Ashby} {et~al.}(2009){Ashby}, {Stern}, {Brodwin}, {Griffith},
  {Eisenhardt}, {Koz{\l}owski}, {Kochanek}, {Bock}, {Borys}, {Brand}, {Brown},
  {Cool}, {Cooray}, {Croft}, {Dey}, {Eisenstein}, {Gonzalez}, {Gorjian},
  {Grogin}, {Ivison}, {Jacob}, {Jannuzi}, {Mainzer}, {Moustakas},
  {R{\"o}ttgering}, {Seymour}, {Smith}, {Stanford}, {Stauffer}, {Sullivan},
  {van Breugel}, {Willner}, \& {Wright}}]{ashby2009}
{Ashby}, M.~L.~N., {Stern}, D., {Brodwin}, M., {Griffith}, R., {Eisenhardt},
  P., {Koz{\l}owski}, S., {Kochanek}, C.~S., {Bock}, J.~J., {Borys}, C.,
  {Brand}, K., {Brown}, M.~J.~I., {Cool}, R., {Cooray}, A., {Croft}, S., {Dey},
  A., {Eisenstein}, D., {Gonzalez}, A.~H., {Gorjian}, V., {Grogin}, N.~A.,
  {Ivison}, R.~J., {Jacob}, J., {Jannuzi}, B.~T., {Mainzer}, A., {Moustakas},
  L.~A., {R{\"o}ttgering}, H.~J.~A., {Seymour}, N., {Smith}, H.~A., {Stanford},
  S.~A., {Stauffer}, J.~R., {Sullivan}, I., {van Breugel}, W., {Willner},
  S.~P., \& {Wright}, E.~L. 2009, \apj, 701, 428

\bibitem[{{Bartelmann}(1995)}]{bartelmann1995}
{Bartelmann}, M. 1995, \aap, 299, 11

\bibitem[{{Bartelmann} {et~al.}(1998){Bartelmann}, {Huss}, {Colberg},
  {Jenkins}, \& {Pearce}}]{bartelmann1998}
{Bartelmann}, M., {Huss}, A., {Colberg}, J.~M., {Jenkins}, A., \& {Pearce},
  F.~R. 1998, \aap, 330, 1

\bibitem[{{Bartelmann} {et~al.}(2003){Bartelmann}, {Meneghetti}, {Perrotta},
  {Baccigalupi}, \& {Moscardini}}]{bartelmann2003}
{Bartelmann}, M., {Meneghetti}, M., {Perrotta}, F., {Baccigalupi}, C., \&
  {Moscardini}, L. 2003, \aap, 409, 449

\bibitem[{{Bertin} \& {Arnouts}(1996)}]{bertin1996}
{Bertin}, E. \& {Arnouts}, S. 1996, \aaps, 117, 393

\bibitem[{{Brodwin} {et~al.}(2012){Brodwin}, {Gonzalez}, {Stanford}, {Stern},
  {Dey}, {Zeimann}, {Eisenhardt}, {Mancone}, Plagge, \&
  Carlstrom}]{brodwin2011}
{Brodwin}, M., {Gonzalez}, A.~H., {Stanford}, S.~A., {Stern}, D., {Dey}, A.,
  {Zeimann}, G., {Eisenhardt}, P.~R., {Mancone}, C., Plagge, T., \& Carlstrom,
  J. 2012, \apjl

\bibitem[{{Bruzual} \& {Charlot}(2003)}]{bruzual2003}
{Bruzual}, G. \& {Charlot}, S. 2003, \mnras, 344, 1000

\bibitem[{{Bullock} {et~al.}(2001){Bullock}, {Kolatt}, {Sigad}, {Somerville},
  {Kravtsov}, {Klypin}, {Primack}, \& {Dekel}}]{bullock2001}
{Bullock}, J.~S., {Kolatt}, T.~S., {Sigad}, Y., {Somerville}, R.~S.,
  {Kravtsov}, A.~V., {Klypin}, A.~A., {Primack}, J.~R., \& {Dekel}, A. 2001,
  \mnras, 321, 559

\bibitem[{{Carlberg} {et~al.}(1996){Carlberg}, {Yee}, {Ellingson}, {Abraham},
  {Gravel}, {Morris}, \& {Pritchet}}]{carlberg1996}
{Carlberg}, R.~G., {Yee}, H.~K.~C., {Ellingson}, E., {Abraham}, R., {Gravel},
  P., {Morris}, S., \& {Pritchet}, C.~J. 1996, \apj, 462, 32

\bibitem[{{Cay{\'o}n} {et~al.}(2011){Cay{\'o}n}, {Gordon}, \&
  {Silk}}]{cayon2011}
{Cay{\'o}n}, L., {Gordon}, C., \& {Silk}, J. 2011, \mnras, 415, 849

\bibitem[{{Chabrier}(2003)}]{chabrier2003}
{Chabrier}, G. 2003, \pasp, 115, 763

\bibitem[{{Coe} {et~al.}(2006){Coe}, {Ben{\'{\i}}tez}, {S{\'a}nchez}, {Jee},
  {Bouwens}, \& {Ford}}]{coe2006}
{Coe}, D., {Ben{\'{\i}}tez}, N., {S{\'a}nchez}, S.~F., {Jee}, M., {Bouwens},
  R., \& {Ford}, H. 2006, \aj, 132, 926

\bibitem[{{Dalal} {et~al.}(2004){Dalal}, {Holder}, \& {Hennawi}}]{dalal2004}
{Dalal}, N., {Holder}, G., \& {Hennawi}, J.~F. 2004, \apj, 609, 50

\bibitem[{{D'Aloisio} \& {Natarajan}(2011)}]{dalioso2011}
{D'Aloisio}, A. \& {Natarajan}, P. 2011, \mnras, 415, 1913

\bibitem[{{Donahue} {et~al.}(1998){Donahue}, {Voit}, {Gioia}, {Lupino},
  {Hughes}, \& {Stocke}}]{donahue1998}
{Donahue}, M., {Voit}, G.~M., {Gioia}, I., {Lupino}, G., {Hughes}, J.~P., \&
  {Stocke}, J.~T. 1998, \apj, 502, 550

\bibitem[{{Eke} {et~al.}(2001){Eke}, {Navarro}, \& {Steinmetz}}]{eke2001}
{Eke}, V.~R., {Navarro}, J.~F., \& {Steinmetz}, M. 2001, \apj, 554, 114

\bibitem[{{Enqvist} {et~al.}(2011){Enqvist}, {Hotchkiss}, \&
  {Taanila}}]{enqvist2011}
{Enqvist}, K., {Hotchkiss}, S., \& {Taanila}, O. 2011, JCAP, 4, 17

\bibitem[{{Ettori} {et~al.}(2010){Ettori}, {Gastaldello}, {Leccardi},
  {Molendi}, {Rossetti}, {Buote}, \& {Meneghetti}}]{ettori2010}
{Ettori}, S., {Gastaldello}, F., {Leccardi}, A., {Molendi}, S., {Rossetti}, M.,
  {Buote}, D., \& {Meneghetti}, M. 2010, \aap, 524, A68

\bibitem[{{Fedeli} {et~al.}(2007){Fedeli}, {Bartelmann}, {Meneghetti}, \&
  {Moscardini}}]{fedeli2007}
{Fedeli}, C., {Bartelmann}, M., {Meneghetti}, M., \& {Moscardini}, L. 2007,
  \aap, 473, 715

\bibitem[{{Fedeli} {et~al.}(2008){Fedeli}, {Bartelmann}, {Meneghetti}, \&
  {Moscardini}}]{fedeli2008}
---. 2008, \aap, 486, 35

\bibitem[{{Fedeli} \& {Berciano Alba}(2009)}]{fedeli2009}
{Fedeli}, C. \& {Berciano Alba}, A. 2009, \aap, 508, 141

\bibitem[{{Fedeli} {et~al.}(2006){Fedeli}, {Meneghetti}, {Bartelmann}, {Dolag},
  \& {Moscardini}}]{fedeli2006}
{Fedeli}, C., {Meneghetti}, M., {Bartelmann}, M., {Dolag}, K., \& {Moscardini},
  L. 2006, \aap, 447, 419

\bibitem[{{Ford} {et~al.}(2003){Ford}, {Clampin}, {Hartig}, {Illingworth},
  {Sirianni}, {Martel}, {Meurer}, {McCann}, {Sullivan}, {Bartko}, {Benitez},
  {Blakeslee}, {Bouwens}, {Broadhurst}, {Brown}, {Burrows}, {Campbell},
  {Cheng}, {Feldman}, {Franx}, {Golimowski}, {Gronwall}, {Kimble}, {Krist},
  {Lesser}, {Magee}, {Miley}, {Postman}, {Rafal}, {Rosati}, {Sparks}, {Tran},
  {Tsvetanov}, {Volmer}, {White}, \& {Woodruff}}]{Ford2003}
{Ford}, H.~C., {Clampin}, M., {Hartig}, G.~F., {Illingworth}, G.~D.,
  {Sirianni}, M., {Martel}, A.~R., {Meurer}, G.~R., {McCann}, W.~J.,
  {Sullivan}, P.~C., {Bartko}, F., {Benitez}, N., {Blakeslee}, J., {Bouwens},
  R., {Broadhurst}, T., {Brown}, R.~A., {Burrows}, C.~J., {Campbell}, D.,
  {Cheng}, E.~S., {Feldman}, P.~D., {Franx}, M., {Golimowski}, D.~A.,
  {Gronwall}, C., {Kimble}, R.~A., {Krist}, J.~E., {Lesser}, M.~P., {Magee},
  D., {Miley}, G., {Postman}, M., {Rafal}, M.~D., {Rosati}, P., {Sparks},
  W.~B., {Tran}, H.~D., {Tsvetanov}, Z.~I., {Volmer}, P., {White}, R.~L., \&
  {Woodruff}, R.~A. 2003, in Presented at the Society of Photo-Optical
  Instrumentation Engineers (SPIE) Conference, Vol. 4854, Future EUV/UV and
  Visible Space Astrophysics Missions and Instrumentation. Edited by J. Chris
  Blades, Oswald H. W. Siegmund. Proceedings of the SPIE, Volume 4854, pp.
  81-94 (2003)., ed. J.~C. {Blades} \& O.~H.~W. {Siegmund}, 81--94

\bibitem[{{Gao} {et~al.}(2008){Gao}, {Navarro}, {Cole}, {Frenk}, {White},
  {Springel}, {Jenkins}, \& {Neto}}]{gao2008}
{Gao}, L., {Navarro}, J.~F., {Cole}, S., {Frenk}, C.~S., {White}, S.~D.~M.,
  {Springel}, V., {Jenkins}, A., \& {Neto}, A.~F. 2008, \mnras, 387, 536

\bibitem[{{Gladders} {et~al.}(2003){Gladders}, {Hoekstra}, {Yee}, {Hall}, \&
  {Barrientos}}]{gladders2003}
{Gladders}, M.~D., {Hoekstra}, H., {Yee}, H.~K.~C., {Hall}, P.~B., \&
  {Barrientos}, L.~F. 2003, \apj, 593, 48

\bibitem[{{Gobat} {et~al.}(2011){Gobat}, {Daddi}, {Onodera}, {Finoguenov},
  {Renzini}, {Arimoto}, {Bouwens}, {Brusa}, {Chary}, {Cimatti}, {Dickinson},
  {Kong}, \& {Mignoli}}]{gobat2011}
{Gobat}, R., {Daddi}, E., {Onodera}, M., {Finoguenov}, A., {Renzini}, A.,
  {Arimoto}, N., {Bouwens}, R., {Brusa}, M., {Chary}, R.-R., {Cimatti}, A.,
  {Dickinson}, M., {Kong}, X., \& {Mignoli}, M. 2011, \aap, 526, A133+

\bibitem[{{Gonzalez} {et~al.}(2011){Gonzalez}, {Bouwens}, {Labbe},
  {Illingworth}, {Oesch}, {Franx}, \& {Magee}}]{vgonzalez2011}
{Gonzalez}, V., {Bouwens}, R., {Labbe}, I., {Illingworth}, G., {Oesch}, P.,
  {Franx}, M., \& {Magee}, D. 2011, ArXiv e-prints

\bibitem[{{Hennawi} {et~al.}(2007){Hennawi}, {Dalal}, {Bode}, \&
  {Ostriker}}]{hennawi2007}
{Hennawi}, J.~F., {Dalal}, N., {Bode}, P., \& {Ostriker}, J.~P. 2007, \apj,
  654, 714

\bibitem[{{Hoyle} {et~al.}(2011){Hoyle}, {Jimenez}, \& {Verde}}]{hoyle2011}
{Hoyle}, B., {Jimenez}, R., \& {Verde}, L. 2011, \prd, 83, 103502

\bibitem[{{Huang} {et~al.}(2009){Huang}, {Morokuma}, {Fakhouri}, {Aldering},
  {Amanullah}, {Barbary}, {Brodwin}, {Connolly}, {Dawson}, {Doi}, {Faccioli},
  {Fadeyev}, {Fruchter}, {Goldhaber}, {Gladders}, {Hennawi}, {Ihara}, {Jee},
  {Kowalski}, {Konishi}, {Lidman}, {Meyers}, {Moustakas}, {Perlmutter},
  {Rubin}, {Schlegel}, {Spadafora}, {Suzuki}, {Takanashi}, \&
  {Yasuda}}]{huang2009}
{Huang}, X., {Morokuma}, T., {Fakhouri}, H.~K., {Aldering}, G., {Amanullah},
  R., {Barbary}, K., {Brodwin}, M., {Connolly}, N.~V., {Dawson}, K.~S., {Doi},
  M., {Faccioli}, L., {Fadeyev}, V., {Fruchter}, A.~S., {Goldhaber}, G.,
  {Gladders}, M.~D., {Hennawi}, J.~F., {Ihara}, Y., {Jee}, M.~J., {Kowalski},
  M., {Konishi}, K., {Lidman}, C., {Meyers}, J., {Moustakas}, L.~A.,
  {Perlmutter}, S., {Rubin}, D., {Schlegel}, D.~J., {Spadafora}, A.~L.,
  {Suzuki}, N., {Takanashi}, N., \& {Yasuda}, N. 2009, \apjl, 707, L12

\bibitem[{{Kimble} {et~al.}(2008){Kimble}, {MacKenty}, {O'Connell}, \&
  {Townsend}}]{kimble2008}
{Kimble}, R.~A., {MacKenty}, J.~W., {O'Connell}, R.~W., \& {Townsend}, J.~A.
  2008, in Presented at the Society of Photo-Optical Instrumentation Engineers
  (SPIE) Conference, Vol. 7010, Society of Photo-Optical Instrumentation
  Engineers (SPIE) Conference Series

\bibitem[{{Komatsu} {et~al.}(2011){Komatsu}, {Smith}, {Dunkley}, {Bennett},
  {Gold}, {Hinshaw}, {Jarosik}, {Larson}, {Nolta}, {Page}, {Spergel},
  {Halpern}, {Hill}, {Kogut}, {Limon}, {Meyer}, {Odegard}, {Tucker}, {Weiland},
  {Wollack}, \& {Wright}}]{komatsu2011}
{Komatsu}, E., {Smith}, K.~M., {Dunkley}, J., {Bennett}, C.~L., {Gold}, B.,
  {Hinshaw}, G., {Jarosik}, N., {Larson}, D., {Nolta}, M.~R., {Page}, L.,
  {Spergel}, D.~N., {Halpern}, M., {Hill}, R.~S., {Kogut}, A., {Limon}, M.,
  {Meyer}, S.~S., {Odegard}, N., {Tucker}, G.~S., {Weiland}, J.~L., {Wollack},
  E., \& {Wright}, E.~L. 2011, \apjs, 192, 18

\bibitem[{{Lee} {et~al.}(2005){Lee}, {Jing}, \& {Suto}}]{lee2005}
{Lee}, J., {Jing}, Y.~P., \& {Suto}, Y. 2005, \apj, 632, 706

\bibitem[{{Li} {et~al.}(2006){Li}, {Mao}, {Jing}, {Mo}, {Gao}, \&
  {Lin}}]{li2006}
{Li}, G.~L., {Mao}, S., {Jing}, Y.~P., {Mo}, H.~J., {Gao}, L., \& {Lin}, W.~P.
  2006, \mnras, 372, L73

\bibitem[{{Mead} {et~al.}(2010){Mead}, {King}, {Sijacki}, {Leonard},
  {Puchwein}, \& {McCarthy}}]{mead2010}
{Mead}, J.~M.~G., {King}, L.~J., {Sijacki}, D., {Leonard}, A., {Puchwein}, E.,
  \& {McCarthy}, I.~G. 2010, \mnras, 406, 434

\bibitem[{{Meneghetti} {et~al.}(2003){Meneghetti}, {Bartelmann}, \&
  {Moscardini}}]{meneghetti2003}
{Meneghetti}, M., {Bartelmann}, M., \& {Moscardini}, L. 2003, \mnras, 340, 105

\bibitem[{{Meneghetti} {et~al.}(2010){Meneghetti}, {Fedeli}, {Pace},
  {Gottl{\"o}ber}, \& {Yepes}}]{meneghetti2010}
{Meneghetti}, M., {Fedeli}, C., {Pace}, F., {Gottl{\"o}ber}, S., \& {Yepes}, G.
  2010, \aap, 519, A90+

\bibitem[{{Meneghetti} {et~al.}(2011){Meneghetti}, {Fedeli}, {Zitrin},
  {Bartelmann}, {Broadhurst}, {Gottl{\"o}ber}, {Moscardini}, \&
  {Yepes}}]{meneghetti2011}
{Meneghetti}, M., {Fedeli}, C., {Zitrin}, A., {Bartelmann}, M., {Broadhurst},
  T., {Gottl{\"o}ber}, S., {Moscardini}, L., \& {Yepes}, G. 2011, \aap, 530,
  A17

\bibitem[{{Meylan} {et~al.}(2006){Meylan}, {Jetzer}, {North}, {Schneider},
  {Kochanek}, \& {Wambsganss}}]{ksw2006}
{Meylan}, G., {Jetzer}, P., {North}, P., {Schneider}, P., {Kochanek}, C.~S., \&
  {Wambsganss}, J., eds. 2006, {Gravitational Lensing: Strong, Weak and Micro},
  ed. G.~{Meylan}, P.~{Jetzer}, P.~{North}, P.~{Schneider}, C.~S. {Kochanek},
  \& J.~{Wambsganss}

\bibitem[{{Mullis} {et~al.}(2005){Mullis}, {Rosati}, {Lamer}, {B{\" o}hringer},
  {Schwope}, {Schuecker}, \& {Fassbender}}]{mullis2005}
{Mullis}, C.~R., {Rosati}, P., {Lamer}, G., {B{\" o}hringer}, H., {Schwope},
  A., {Schuecker}, P., \& {Fassbender}, R. 2005, \apjl, 623, L85

\bibitem[{{Navarro} {et~al.}(1996){Navarro}, {Frenk}, \& {White}}]{nfw1996}
{Navarro}, J.~F., {Frenk}, C.~S., \& {White}, S.~D.~M. 1996, \apj, 462, 563

\bibitem[{{Navarro} {et~al.}(1997){Navarro}, {Frenk}, \& {White}}]{nfw1997}
---. 1997, \apj, 490, 493

\bibitem[{{Oguri} {et~al.}(2012){Oguri}, {Bayliss}, {Dahle}, {Sharon},
  {Gladders}, {Natarajan}, {Hennawi}, \& {Koester}}]{oguri2011}
{Oguri}, M., {Bayliss}, M.~B., {Dahle}, H., {Sharon}, K., {Gladders}, M.~D.,
  {Natarajan}, P., {Hennawi}, J.~F., \& {Koester}, B.~P. 2012, \mnras, 2189

\bibitem[{{Oguri} {et~al.}(2009){Oguri}, {Hennawi}, {Gladders}, {Dahle},
  {Natarajan}, {Dalal}, {Koester}, {Sharon}, \& {Bayliss}}]{oguri2009}
{Oguri}, M., {Hennawi}, J.~F., {Gladders}, M.~D., {Dahle}, H., {Natarajan}, P.,
  {Dalal}, N., {Koester}, B.~P., {Sharon}, K., \& {Bayliss}, M. 2009, \apj,
  699, 1038

\bibitem[{{Press} \& {Schechter}(1974)}]{press1974}
{Press}, W.~H. \& {Schechter}, P. 1974, \apj, 187, 425

\bibitem[{{Puchwein} \& {Hilbert}(2009)}]{puchwein2009}
{Puchwein}, E. \& {Hilbert}, S. 2009, \mnras, 398, 1298

\bibitem[{{Reddy} {et~al.}(2008){Reddy}, {Steidel}, {Pettini}, {Adelberger},
  {Shapley}, {Erb}, \& {Dickinson}}]{reddy2008}
{Reddy}, N.~A., {Steidel}, C.~C., {Pettini}, M., {Adelberger}, K.~L.,
  {Shapley}, A.~E., {Erb}, D.~K., \& {Dickinson}, M. 2008, \apjs, 175, 48

\bibitem[{{Rosati} {et~al.}(2009){Rosati}, {Tozzi}, {Gobat}, {Santos},
  {Nonino}, {Demarco}, {Lidman}, {Mullis}, {Strazzullo}, {B{\"o}hringer},
  {Fassbender}, {Dawson}, {Tanaka}, {Jee}, {Ford}, {Lamer}, \&
  {Schwope}}]{rosati2009}
{Rosati}, P., {Tozzi}, P., {Gobat}, R., {Santos}, J.~S., {Nonino}, M.,
  {Demarco}, R., {Lidman}, C., {Mullis}, C.~R., {Strazzullo}, V.,
  {B{\"o}hringer}, H., {Fassbender}, R., {Dawson}, K., {Tanaka}, M., {Jee}, J.,
  {Ford}, H., {Lamer}, G., \& {Schwope}, A. 2009, \aap, 508, 583

\bibitem[{{Sanderson} {et~al.}(2009){Sanderson}, {Edge}, \&
  {Smith}}]{sanderson2009}
{Sanderson}, A.~J.~R., {Edge}, A.~C., \& {Smith}, G.~P. 2009, \mnras, 398, 1698

\bibitem[{{Schmidt} \& {Allen}(2007)}]{schmidt2007}
{Schmidt}, R.~W. \& {Allen}, S.~W. 2007, \mnras, 379, 209

\bibitem[{{Stanford} {et~al.}(2012){Stanford}, {Brodwin}, {Gonzalez}, {Stern},
  {Dey}, {Zeimann}, {Eisenhardt}, \& {Mancone}}]{stanford2011}
{Stanford}, S.~A., {Brodwin}, M., {Gonzalez}, A.~H., {Stern}, D., {Dey}, A.,
  {Zeimann}, G., {Eisenhardt}, P.~R., \& {Mancone}, C. 2012, \apjl, 000, 000

\bibitem[{{Tinker} {et~al.}(2008){Tinker}, {Kravtsov}, {Klypin}, {Abazajian},
  {Warren}, {Yepes}, {Gottl{\"o}ber}, \& {Holz}}]{tinker2008}
{Tinker}, J., {Kravtsov}, A.~V., {Klypin}, A., {Abazajian}, K., {Warren}, M.,
  {Yepes}, G., {Gottl{\"o}ber}, S., \& {Holz}, D.~E. 2008, \apj, 688, 709

\bibitem[{{Wambsganss} {et~al.}(2004){Wambsganss}, {Bode}, \&
  {Ostriker}}]{wambsganss2004}
{Wambsganss}, J., {Bode}, P., \& {Ostriker}, J.~P. 2004, \apjl, 606, L93

\bibitem[{{Wambsganss} {et~al.}(2008){Wambsganss}, {Ostriker}, \&
  {Bode}}]{wambsganss2008}
{Wambsganss}, J., {Ostriker}, J.~P., \& {Bode}, P. 2008, \apj, 676, 753

\bibitem[{{Williamson} {et~al.}(2011){Williamson}, {Benson}, {High},
  {Vanderlinde}, {Ade}, {Aird}, {Andersson}, {Armstrong}, {Ashby}, {Bautz},
  {Bazin}, {Bertin}, {Bleem}, {Bonamente}, {Brodwin}, {Carlstrom}, {Chang},
  {Chapman}, {Clocchiatti}, {Crawford}, {Crites}, {de Haan}, {Desai}, {Dobbs},
  {Dudley}, {Fazio}, {Foley}, {Forman}, {Garmire}, {George}, {Gladders},
  {Gonzalez}, {Halverson}, {Holder}, {Holzapfel}, {Hoover}, {Hrubes}, {Jones},
  {Joy}, {Keisler}, {Knox}, {Lee}, {Leitch}, {Lueker}, {Luong-Van}, {Marrone},
  {McMahon}, {Mehl}, {Meyer}, {Mohr}, {Montroy}, {Murray}, {Padin}, {Plagge},
  {Pryke}, {Reichardt}, {Rest}, {Ruel}, {Ruhl}, {Saliwanchik}, {Saro},
  {Schaffer}, {Shaw}, {Shirokoff}, {Song}, {Spieler}, {Stalder}, {Stanford},
  {Staniszewski}, {Stark}, {Story}, {Stubbs}, {Vieira}, {Vikhlinin}, \&
  {Zenteno}}]{williamson2011}
{Williamson}, R., {Benson}, B.~A., {High}, F.~W., {Vanderlinde}, K., {Ade},
  P.~A.~R., {Aird}, K.~A., {Andersson}, K., {Armstrong}, R., {Ashby}, M.~L.~N.,
  {Bautz}, M., {Bazin}, G., {Bertin}, E., {Bleem}, L.~E., {Bonamente}, M.,
  {Brodwin}, M., {Carlstrom}, J.~E., {Chang}, C.~L., {Chapman}, S.~C.,
  {Clocchiatti}, A., {Crawford}, T.~M., {Crites}, A.~T., {de Haan}, T.,
  {Desai}, S., {Dobbs}, M.~A., {Dudley}, J.~P., {Fazio}, G.~G., {Foley}, R.~J.,
  {Forman}, W.~R., {Garmire}, G., {George}, E.~M., {Gladders}, M.~D.,
  {Gonzalez}, A.~H., {Halverson}, N.~W., {Holder}, G.~P., {Holzapfel}, W.~L.,
  {Hoover}, S., {Hrubes}, J.~D., {Jones}, C., {Joy}, M., {Keisler}, R., {Knox},
  L., {Lee}, A.~T., {Leitch}, E.~M., {Lueker}, M., {Luong-Van}, D., {Marrone},
  D.~P., {McMahon}, J.~J., {Mehl}, J., {Meyer}, S.~S., {Mohr}, J.~J.,
  {Montroy}, T.~E., {Murray}, S.~S., {Padin}, S., {Plagge}, T., {Pryke}, C.,
  {Reichardt}, C.~L., {Rest}, A., {Ruel}, J., {Ruhl}, J.~E., {Saliwanchik},
  B.~R., {Saro}, A., {Schaffer}, K.~K., {Shaw}, L., {Shirokoff}, E., {Song},
  J., {Spieler}, H.~G., {Stalder}, B., {Stanford}, S.~A., {Staniszewski}, Z.,
  {Stark}, A.~A., {Story}, K., {Stubbs}, C.~W., {Vieira}, J.~D., {Vikhlinin},
  A., \& {Zenteno}, A. 2011, \apj, 738, 139

\bibitem[{{Wu} \& {Hammer}(1993)}]{wu1993}
{Wu}, X.-P. \& {Hammer}, F. 1993, \mnras, 262, 187

\bibitem[{{Zwicky}(1933)}]{zwicky1933}
{Zwicky}, F. 1933, Helvetica Physica Acta, 6, 110

\end{thebibliography}

\end{document}